\documentclass[aip,amsmath,amssymb, reprint]{revtex4-1}
\usepackage{physics}
\usepackage{amsmath}
\usepackage{mathptmx,bm}
\usepackage{graphicx}
\usepackage{dcolumn}
\usepackage{tabularx}
\usepackage{epsf}
\usepackage{color}
\usepackage{hyperref}
\hypersetup{breaklinks,colorlinks,linkcolor=blue,citecolor=blue,urlcolor=blue}

\usepackage{epstopdf}%
\usepackage{verbatim}
\usepackage[english]{babel}
\newcommand{\bea}{\begin{eqnarray}}
	\newcommand{\eea}{\end{eqnarray}}
\usepackage{amsfonts}
\usepackage{graphicx}
\usepackage{dcolumn}
\usepackage{bm}
\usepackage[utf8]{inputenc}
\usepackage[T1]{fontenc}
\usepackage{mathptmx}
\usepackage{etoolbox}

\makeatletter
\def\@email#1#2{%
	\endgroup
	\patchcmd{\titleblock@produce}
	{\frontmatter@RRAPformat}
	{\frontmatter@RRAPformat{\produce@RRAP{*#1\href{mailto:#2}{#2}}}\frontmatter@RRAPformat}
	{}{}
}%
\makeatother

\begin{document}

\title{Thermal transport in fullerene-based molecular junctions: Molecular dynamics simulations}

\author{Joanna Li}
\affiliation{Department of Physics, University of Toronto, 60 Saint George St., Toronto, Ontario M5S 1A7, Canada}
\affiliation{Division of Engineering Science, University of Toronto, 42 Saint George St., Toronto, Ontario M5S 2E4, Canada}

\author{Jonathan J. Wang}
\affiliation{Chemical Physics Theory Group, Department of Chemistry, University of Toronto, 80 Saint George St., Toronto, Ontario M5S 3H6, Canada}

\author{Dvira Segal}
\affiliation{Chemical Physics Theory Group, Department of Chemistry, University of Toronto, 80 Saint George St., Toronto, Ontario M5S 3H6, Canada}
\affiliation{Department of Physics, University of Toronto, 60 Saint George St., Toronto, Ontario M5S 1A7, Canada}

\email{dvira.segal@utoronto.ca}

\date{\today}

\begin{abstract}
We investigate phonon thermal transport of fullerene-based single-molecule junctions by employing classical molecular dynamics simulations. 
The thermal conductances of fullerene monomers, dimers, and trimers are computed through three distinct molecular dynamics methods, by following the equilibration dynamics in one method, and using two other nonequilibrium simulation methods. We discuss technical aspects of the simulation techniques, and show that their predictions for the thermal conductance agree. 
Our simulations reveal that the thermal conductance of fullerene monomer and dimer junctions are similar, yet the thermal conductance of trimer junctions is significantly reduced. 
This study could assist in the design of high-performing thermoelectric junctions, where low thermal conductance is desired.  
\end{abstract}

\maketitle

\section{Introduction}

Recent groundbreaking experiments reported the thermal conductance of selected single molecules in the configuration of a metal-molecule-metal junction \cite{CuiExp19,GotsmannExp19,GotsmannExp23}. 
These efforts are of an interest from the fundamental side for understanding thermal transport mechanisms at the nanoscale level of a single molecule \cite{Baowen12,Leitner15,RevA,RevG}, and for applications in electronics, energy conversion, and novel materials development \cite{Pop10,Rev14,PramodR,Yoon20,BaowenR21}.
In particular, for flexible organic-based electronic devices, it is important to
 identify molecules that exhibit large thermal conductance, beneficial for transferring
 away dissipated heat and maintaining the integrity of the junction. 
Conversely, thermoelectric devices show improved efficiency when reducing their thermal transport, including the electronic, phononic and photonic components. 

Single-molecule thermal transport measurements, as reported in Refs. \citenum{CuiExp19,GotsmannExp19,GotsmannExp23}
access the {\it total} thermal conductance of the junction, but they
cannot distinguish between its electronic, phononic or photonic contributions. 
Since alkane chains are known to be poor conductors of electrons \cite{alkaneR}, their thermal conductance is assumed to be governed by phonons. As such, alkane molecules became prominent testbeds for estimating phononic conduction in molecular junctions.
These molecules were examined in both single-molecule  \cite{CuiExp19,GotsmannExp19} and self-assembled \cite{Wang06,Dlott07,Cahill12,GotsmannExp14,Shub15,Shub17} experiments. Alkane chains were also heavily studied theoretically and computationally  \cite{Dvira2003,Pawel11,Pauly16,Pauly18,Nitzan20,Lu2021,Nitzan22,JW-HeatMD, JW-fluct} with the objective
to explore mechanisms of thermal transport, develop 
methodologies,
and benchmark theoretical-computational predictions.

In fact, due to the challenge in measuring thermal transport of single-molecule junctions, simulations play a leading role in guiding the search for effective thermal transport junctions \cite{Luo13}.
Two distinct computational tools dominate this field:
(i) Employing a quantum method that 
integrates density functional theory (DFT) with the
nonequilibrium Green’s function (NEGF) technique \cite{DFTNEGF1,Pauly16}. This
framework does not take explicit many-body interactions into account. It enables separate computations of the contributions of electrons 
or phonons to the thermal conductance.
The DFT-NEGF method builds on the Landauer equation, and it assumes noninteracting electrons and harmonic interactions in the junction, ignoring dynamical effects: 
Simulations are executed on a frozen configuration, yet frequencies and metal-molecule hybridization energies are obtained from DFT calculations.
(ii) Using classical molecular dynamics (MD) simulations that build on parametrized force fields. 
Conventional classical MD simulations cannot provide any information on electronic thermal transport. However, for the vibrational-phononic component, MD simulations offer great flexibility as such methods are not restricted to the harmonic force field. Furthermore, MD simulations naturally allow the molecule and the metal atoms to move throughout the simulation, thus taking into account dynamical effects. 
However, in the standard implementation of MD simulations, such as that used in the LAMMPS simulator \cite{LAMMPS}, potentials are phenomenological with fixed parametrization. 
In our study here we employ such an MD approach with
details given in Sec. \ref{Method-sec}.

Fullerenes, carbon-based cage-like molecules, support a wide range of applications in nanoscience in e.g., drug delivery, catalysis, organic photovoltaics, and energy storage devices \cite{Fullerene-rev}. 
In particular, electrical conductance and thermopower measurements of C60 monomer and dimers
were carried out in Ref. \citenum{Evangeli}. It was further suggested in that study 
that stacks of C60 molecules could be candidates for thermoelectric applications by achieving high efficiency.
However, a DFT-NEGF computational study of thermal transport in monomer and dimer fullerenes did not support this proposal; it was shown in Ref. \citenum{Klockner} that phonons dominate and limit the
thermoelectric performance of dimer fullerene junctions.
 
\begin{figure*}[htbp]
    \centering
    \includegraphics[width=1.5\columnwidth]{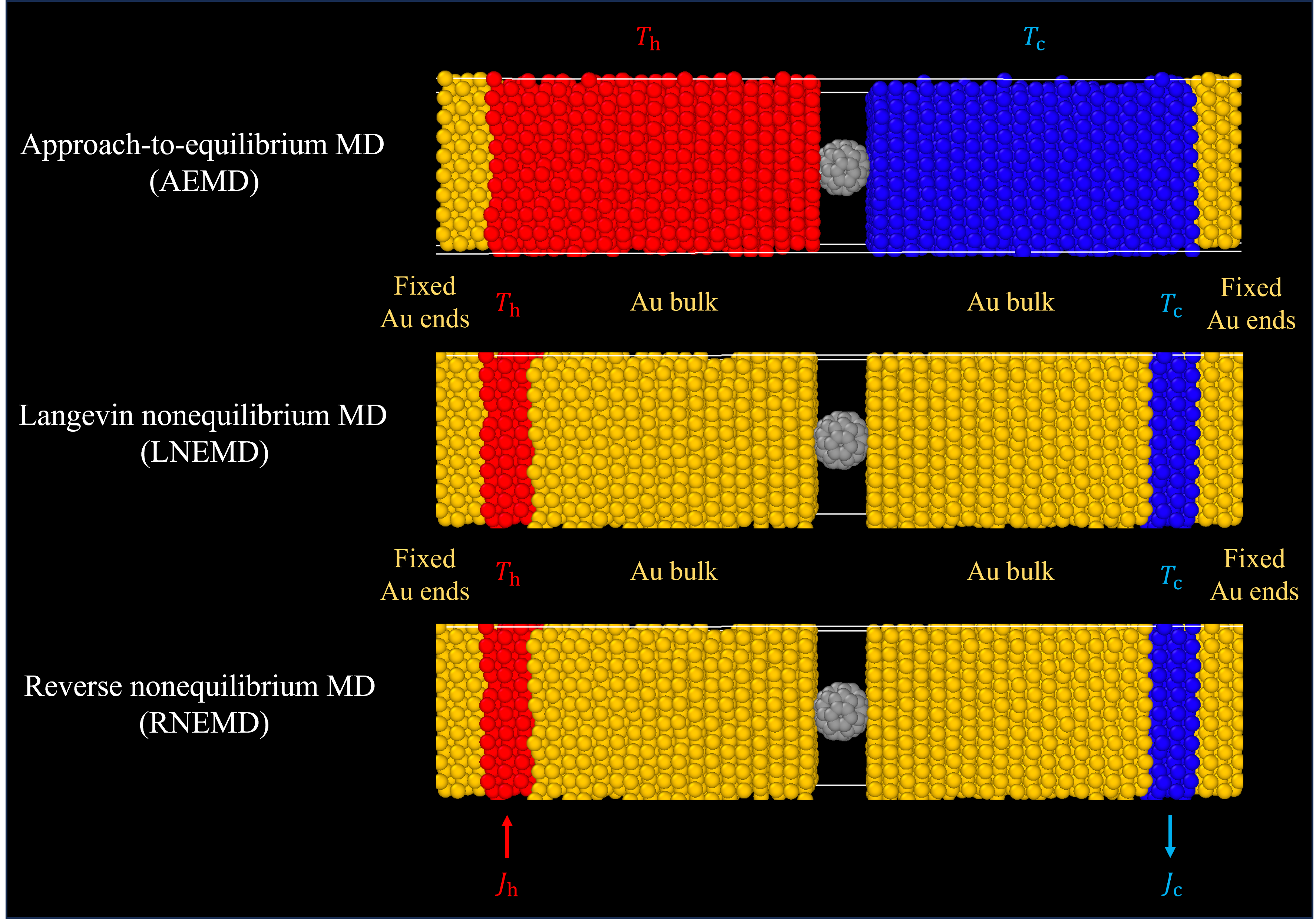}
\caption{Visualization of a single-molecule gold-fullerene-gold nanojunction (OVITO \cite{OVITO}) and the methods used to simulate thermal transport as described in Sec. \ref{Method-sec}.
(Top)  In the AEMD method, the two metals are initially separately thermalized to $T_h$ and $T_c$,    while the molecule is prepared at the average temperature. The thermal conductance is evaluated from the equilibration dynamics, see Sec. \ref{sec:AEMD}.
(Centre) In the LNEMD method, Langevin thermostats are attached to a certain region at the back of the metals away from the junction, thermalizing them to temperatures $T_h$ (left) and $T_c$ (right). The heat current is evaluated by calculating the net heat exchange between the thermostats and the metal at each contact, see Sec. \ref{sec:LNEMD}.
(Bottom) In the RNEMD method, the heat current into the system (and equivalently out) is imposed as a parameter, resulting in a temperature drop on the junction. The method is described in Sec. \ref{sec:RNEMD}.
In all three approaches, a group of 270 gold atoms at the left and right boundaries (colored in yellow) are fixed, to keep the structure from collapsing. 
    }
    \label{Fig1}
\end{figure*}

Motivated by thermopower measurements of fullerene monomers and dimers metal-molecule-metal junctions,
and the study of their electronic, thermal and thermoelectric properties using the DFT-NEGF approach, 
we simulate here the phononic thermal conductance of fullerene junctions using the complementary classical MD approach. Contrary to the Landauer equation, classical MD allows us to include anharmonic interactions and dynamical effects.
Our first main objective in this study is the computation of the thermal conductance of fullerenes in a junction configuration, including monomers, dimers and trimers junctions. 
Concurrently, our second goal is to continue establishing molecular dynamics simulation methods for the study of single-molecule thermal transport. We achieve this goal by adopting and comparing three molecular dynamics approaches for thermal transport calculations.
In this respect, while in our current calculations we utilize parameterized force field potentials and rely on fully classical simulations, more advanced studies could benefit from the adoption of first-principle machine learning potentials and the integration of quantum corrections. The confidence established in simulation methods through this study will serve future investigations that aim to employ more accurate potentials and incorporate quantum corrections.

The plan of this paper is as follows. In Sec. \ref{Method-sec}, we describe three  molecular dynamics methods for the simulation of thermal transport in single molecule junctions: The Approach to equilibrium MD, which follows  equilibration dynamics,
 the Langevin nonequilibrium MD, in which we thermostat the end sections of the two metals to different temperatures, and compute the net heat current in the system, and 
 the reversed nonequilibrium MD technique, where we do the opposite: We impose the net heat current and compute the temperature profile.
Simulation results are presented in Sec. \ref{results-sec}, where we discuss the heat current trend with temperature difference and the number of stacked fullerenes.
We summarize our work in Sec. \ref{Sum-sec} offering future perspectives.

\section{Setup and Simulation Techniques}
\label{Method-sec}

We construct metal-molecule-metal junctions and perform thermal transport simulations using three methods: Approach to Equilibrium Molecular Dynamics (AEMD), Langevin Non Equilibrium Molecular Dynamics (LNEMD), and the Reversed Non Equilibrium Molecular Dynamics (RNEMD), as illustrated in Fig. \ref{Fig1}. We calculate the thermal conductance of the junction $G$ in the three methods. In addition, the latter two nonequilibrium methods enable us to study the heat current in the steady state limit as a function of factors such as the imposed temperature difference. 

The junction consists of two gold leads, each containing 2160 atoms. At the ends of each lead, the Au atoms are fixed in order to hold the junction and prevent the two pieces of gold from collapsing together. These atoms cannot move in the simulation. Periodic boundaries are implemented in the $x$ and $y$ dimensions (planes of gold), but the simulation box is extended in the $z$ dimension; the top of the box is placed farther away from the atoms. 

Atom types comprise of Au in the gold leads and C in the fullerene molecule placed between the two leads, see Fig. \ref{Fig1}. The force fields used are the same for all three methods. Interactions within the fullerene molecule are modelled using an adaptive intermolecular reactive bond order (AIREBO) potential that is optimized for carbon-carbon interactions \cite{AIREBO}. Au-Au interactions are given by the embedded atom model. A Lennard-Jones potential is set for long-range, non-bonded interactions between all atoms. The parameters for the potentials are summarized in Table \ref{tab:Parameters}.

\begin{table*}[htbp]
    \begin{center}
    \caption{Parameters and potentials used in the MD simulations}
	\begin{tabular}{|p{4cm}|p{2cm}|p{8cm}|} 
 \hline
\bf Potential & \bf Interaction & \bf Parameters \\
 \hline
Lennard-Jones & Au-Au \newline \newline Au-C & $ \epsilon = 0.00169$ eV, $\sigma = 2.935$ \AA, $r_{cut} = 10 $ \AA \cite{JW-HeatMD} \newline \newline $ \epsilon = 0.001273$ eV, $\sigma = 2.9943$ \AA, $r_{cut} = 10 $ \AA \cite{Pishkenari} \\
 \hline
Embedded Atom Model \newline (EAM/Alloy) & Au-Au & Ref. \citenum{Grochola} \\ 
 \hline
AIREBO  \newline (Lennard-Jones \newline REBO and torsion terms) & C-C &   $ \epsilon = 0.0028$ eV, $\sigma = 3.4$ \AA, $r_{cut} = 8.5 $ \AA 
\cite{Pishkenari} \newline 
Ref. \citenum{AIREBO}
\\
 \hline
\end{tabular}
\label{tab:Parameters}
\end{center}
\end{table*}

\begin{figure*}[htpb]
    \centering
    \includegraphics[width=2\columnwidth]{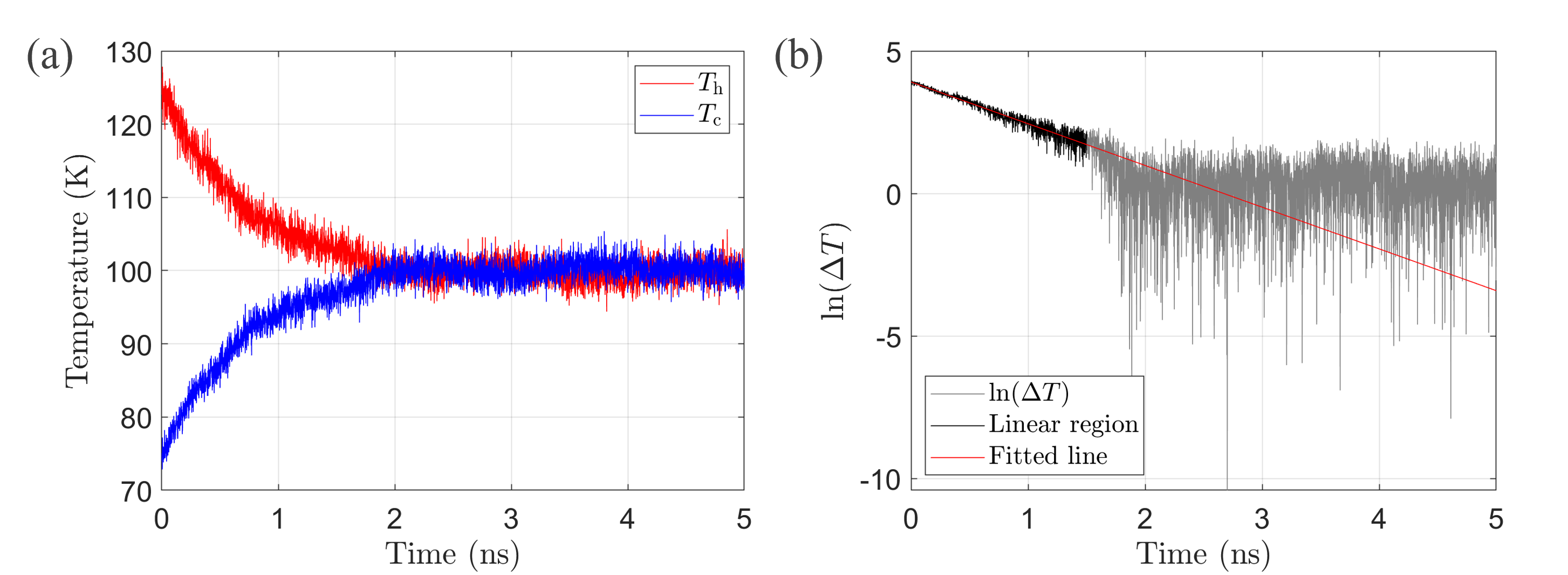}
    \caption{Example of AEMD simulations of thermal conductance of C60 monomer nanojunctions.
(a) The two gold metals are prepared initially at two different temperatures $T_h$ (red) and $T_c$ (blue), and these temperatures are monitored over time once the metals are put into contact through the fullerene molecule.
(b) Data analysis of an AEMD simulation.
A plot of $\ln(\Delta T(t))$ over time provides the decay timescale $\tau$, which is used for the determination of the thermal conductance of the junction through Eq. (\ref{eq:DTsol}). The linear fit is performed over data up to 1.5 ns (black), while beyond that (gray) the system reaches thermal equilibrium.
}
\label{Fig2}
\end{figure*}
\subsection{Approach to equilibrium MD}
\label{sec:AEMD}

The approach to equilibrium molecular dynamics 
was described in several studies, for example Refs. \citenum{Keblinski-AEMD,Martin,JW-HeatMD}. In this method, the thermal conductance is calculated from the equilibration dynamics by monitoring the temperatures of the two bulk solids. 
The method relies on the phenomenological Newton's law of cooling. We start with a nonequilibrium condition where the two separate metals are prepared at distinct temperatures. Once the metals are attached via the fullerene at the junction, energy is exchanged between the metals and the system eventually reaches a global equilibrium state. By monitoring the temperature of the two bulk metals as they equilibrate, one can determine the thermal conductance of the junction as follows:
Newton's cooling law for each metal is given by
\bea
C\frac{dT_h(t)}{dt} &=& -G\left[T_h(t)-T_c(t)\right ],
\nonumber\\
C\frac{dT_c(t)}{dt} &=& -G\left[T_c(t)-T_h(t)\right].
\label{eq:New}
\eea
Here, the temperatures change in time towards equilibrium. 
$C$ is the heat capacity of the the metal. Using the Dulong–Petit law, it is approximated as $C=3k_BN_{\text{Au}}$ with $N$ the number of moving gold atoms in that metal; $k_B$ is the Boltzmann's constant. Defining the temperature difference, $\Delta T(t)=T_h(t)-T_c(t)$, the two equations (\ref{eq:New}) can be combined,
\bea
\frac{d\Delta T(t)}{dt} = -\frac{2G}{C} \Delta T(t).
\eea
The solution to this first-order differential equation is
\bea 
\Delta T(t) = \Delta T(0)\; e^{-t/\tau},
\label{eq:DTsol}
\eea
where $\tau\equiv C/2G$ is the equilibration timescale. We also define the averaged temperature as $\bar T(t)= [T_h(t)+ T_c(t)]/2$. This temperature is evaluated as the average over the full bulk.

\subsubsection{Procedure}
In the AEMD method, NPT equilibration is first carried out to relax the system and the simulation box, barostatting for zero pressure and thermostatting to a target temperature, $\bar{T}$. Subsequent NVT equilibration is carried out separately on the two Au leads to achieve the starting temperatures, $T_h(0)$ and $T_c(0)$. The fullerene molecules at the center are equilibrated to the averaged value $\bar T$. These 
equilibrations are performed over 1.5 ns with 1 fs timesteps. As for the production run: An NVE simulation is conducted for 5 ns, so the temperatures of the Au relax to the equilibrium value. Given the symmetry of the setup, the temperature of each metal, and the fullerenes, eventually reaches
$\bar{T}= [T_h(0)+ T_c(0)]/2$. Temperature data is logged every 1000 steps of 1 fs timesteps, equivalent to every 1 ps.
\subsubsection{Data Analysis}

We display an example of AEMD simulations in Fig.~\ref{Fig2}
with initial temperatures of 125 K and 75 K at the two metals. 
In panel (a), we follow the equilibration dynamics, which takes place within about 2 ns after contact. After that time, the temperatures of the two metals reach the averaged value (100 K), showing some temporal fluctuations.
Temperatures are evaluated from the kinetic energy assuming equipartition. 
The temperature of each metal is calculated as the average of the bulk gold (1890 atoms).

To extract the relaxation timescale as described in Eq. (\ref{eq:DTsol}), we plot 
in Fig.~\ref{Fig2} (b) the evolving  temperature difference as a function of time, illustrating an exponential decay within the first $\approx$2 ns. In fact, we only use the data marked in black (up to 1.5 ns) for the calculation of the thermal conductance, with results summarized in Fig. \ref{Fig4}. We repeat this calculation at two different averaged temperatures, $\bar{T}=100$ K and $\bar{T}=300$ K, with $\Delta T = 50$ K and $\Delta T = 100$ K respectively, in both cases reaching $G\approx 50$ pW/K.

The primary advantage of this method lies in its relatively short simulation times. However, the relaxation dynamics may deviate from Newton's cooling law, displaying multiple relaxation times (as illustrated below in Fig. \ref{Fig7}). While offering insights into nanoscale dynamics, these nontrivial trends pose challenges in calculating the thermal conductance of the junction.

\subsection{Langevin nonequilibrium MD}
\label{sec:LNEMD}
In the Langevin Non-Equilibrium MD approach,
several layers of gold atoms at each metal are continuously thermostated to a given temperature, which is different at the two ends, thus generating a nonequilibrium steady state, see Fig. \ref{Fig1}(b). 
Examples of recent studies adopting this method for thermal transport calculations in polymer junctions include Refs. \citenum{Nitzan20,Lu2021,Nitzan22}. 

We use Langevin thermostats characterized by their temperature, $T_h$ and $T_c$, with the damping timescale $\gamma^{-1}$. Here, $\gamma$ corresponds to the rate of thermalization with the local Langevin bath. 
The instantaneous heat current at each contact can be evaluated from the Langevin dynamics, by building the trajectory $\Delta E_{\text {cold}}(t)$ and $\Delta E_{\text {hot}}(t)$ as the cumulative heat exchanged from time $t=0$ to time $t$.
Our sign convention is such that heat input from the 
 system's atoms to the attached thermostats is defined as positive, so 
$\Delta E_{\text {cold}}(t)$ is positive while $\Delta E_{\text {hot}}(t)$ is negative.

\subsubsection{Procedure}
In the LNEMD method, a temperature bias is imposed on the two gold leads using Langevin baths. NPT equilibration is first carried out to relax the system and simulation box, barostatting for zero pressure and thermostatting to a target temperature, $\bar{T}$. A subsequent NVT equilibration is carried out by equilibrating the whole system to the same target temperature. Each equilibration is performed over 1.5 ns with 1 fs timesteps. As for the production step, NVE simulations are performed to produce steady-state temperature profiles. At this stage, Langevin thermostats are applied to the two gold leads, one set at a high temperature $T_h$ and the other at a cold temperature $T_c$. The damping rate $\gamma^{-1}$ is set to 0.05 ps, which was demonstrated in Ref. \citenum{JW-fluct} to minimally interfere with the junction's dynamics. These production runs are executed for a total of 10 ns, with results averaged over the last 9 ns. Temperature data is logged every 1000 steps of 1 fs timesteps, equivalent to every 1 ps.

\begin{figure}[htbp]
\centering
\includegraphics[width=\columnwidth]{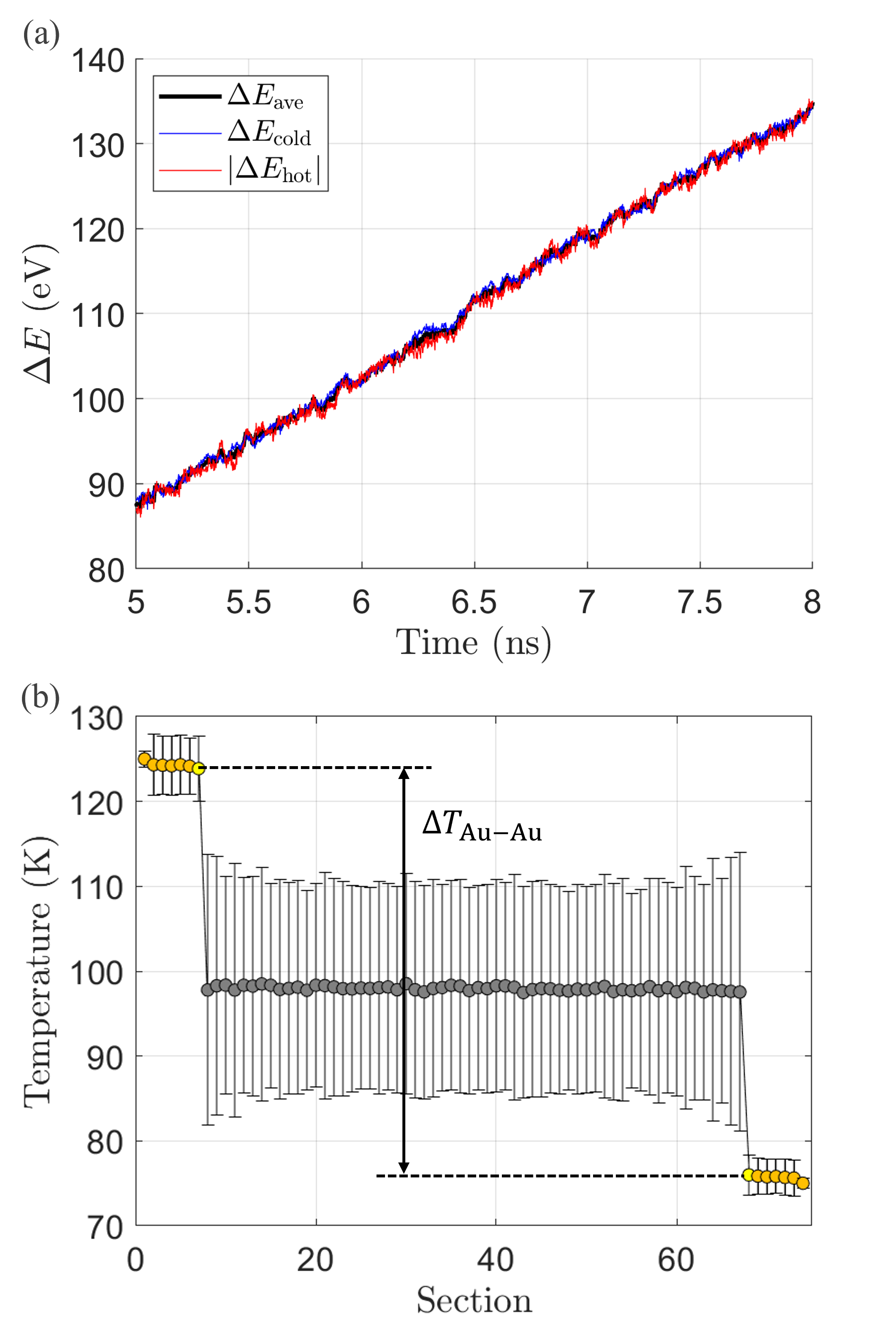}
\caption{Data collected in the LNEMD method.
 (a) Cumulative energy exchange of each metal with the
Langevin thermostats. 
$\Delta E_{\text {hot}}$ and $\Delta E_{\text {cold}}$ are the cumulative heat absorbed and emitted by Au atoms connected to the hot and cold thermostats, respectively. 
$\Delta E_{\text {ave}} (t)$ is the average value of these two trajectories (maintaining the sign positive for heat entering the thermostats). 
(b) Temperature profile across the junction in the steady state limit. Sections correspond to 270 Au atoms and single C atoms in the fullerene. 
Simulation parameters are $T_h = 125$ K, $T_{c}=75$ K, and
$\gamma^{-1} = 0.05$ ps.
    }
    \label{Fig3}
\end{figure}
\subsubsection{Data Analysis}
Fig. \ref{Fig3} exemplifies data collection and analysis in the LNEMD method. Raw data of heat exchange with the thermostats is displayed in Fig. \ref{Fig3}(a). The amount of heat absorbed by the gold atoms attached to the hot baths (red) is equal to the amount of heat released to the cold thermostats (blue). 
In the nonequilibrium steady state, the junction responds with the development of a temperature bias internally on the junction.
In Fig. \ref{Fig3}(b) we display this temperature profile. 
The temperature of each carbon atom is evaluated from its kinetic energy, and as expected, this measure for temperature shows large fluctuations (see error bars). For the gold metal, we evaluate temperature for slabs with 270 atoms.
The temperature drop $\Delta T_{\text {Au-Au}}$ is further marked in Fig. \ref{Fig3}(b), evaluated at the edge gold atoms. 
The heat current in the junction is calculated from the 
slope in panel (a) as
$J=(|\Delta E(t_2)| -|\Delta E(t_1)| )/(t_2-t_1)$, 
with $t_2>t_1$ and for a long time interval. The thermal conductance is given by the ratio $G=J/\Delta T_{\text {Au-Au}}$.
As can be seen in Fig. \ref{Fig3}(b), the thermal resistance is dominated by the metal-molecule interface \cite{ITR,ITR2}. To enhance thermal conductance, this boundary effect could be alleviated by e.g., using other materials as bulk contacts.

The LNEMD method is convenient to use, specifically if one is interested in studying the heat current-temperature difference trends e.g., towards studying the thermal diode effect and in investigations of the heat current fluctuations \cite{JW-fluct}.  One should be aware however of two issues:
(i) Properties of the molecular junctions such as its thermal conductance should not depend on the damping rate $\gamma$, and this should be verified for each system. This simulation parameter corresponds to the Au-thermostat interaction, and it is not part of the atomic model.
(ii) Standard simulators such as LAMMPS do not implement solvers for stochastic differential equations, leading to violations of e.g., heat exchange fluctuation relations  \cite{JW-fluct}.

\subsection{Reversed nonequilibrium MD}
\label{sec:RNEMD}


In the RNEMD method, we use a reversed procedure for determining the current-thermal bias relationship. 
In this approach, the magnitude of the heat current is imposed as a parameter with $J_h$ energy per unit time added on average to one side---and the same amount extracted from the opposite metal contact. 
When simulating for a long enough time, the system reaches a nonequilibrium steady state building
a temperature bias in the junction in response to the 
input heat current.
The method was used extensively in recent molecular dynamics studies, even when combined with quantum corrections. Some recent examples include Refs.\citenum{Lloyd, Keblinski,Kiku, Chen,JW-HeatMD}.

\subsubsection{Procedure}
In the RNEMD method, heat is inputted and outputted close to the ends of the two gold leads, see Fig. \ref{Fig1}(c). NPT equilibration is first carried out to relax the system and the simulation box, barostatting for zero pressure and thermostatting to a target temperature, $\bar{T}$. Subsequent NVT equilibration is carried out by equilibrating the whole system to the same target temperature. Each equilibration is performed over 1.5 ns with 1 fs timesteps. In the production step, NVE simulations are performed to produce steady-state temperature profiles. Heat is inputted to one lead's bath and extracted from the other (as kinetic energy per unit time to the atoms), at the same rates. These production runs are executed for a total of 22.5 ns, with results averaged over the last 20 ns. Temperature data is logged every 1000 steps of 1 fs timesteps, equivalent to every 1 ps.

\subsubsection{Data Analysis}
In the steady state limit, we observe that in response to the enacted heat current, a temperature bias develops on the junction, similar to the profile shown in Fig. \ref{Fig3}(b) for the LNEMD method. 
We define the resulting temperature bias on the junction as  $\Delta T_{\text{Au-Au}}$; the temperatures are evaluated on the metal at the proximity of the molecule. This thermal bias allows the calculation of the thermal conductance $G$ using the linear response expression,
$G = \frac{J}{\Delta T}$.
Altogether, in the ``reversed" approach we identify the 
$J$-$\Delta T$ (current-temperature bias) relationship by setting the current and extracting the thermodynamic force, opposite to what is done in the direct, LNEMD method. 

The main advantage of the RNEMD is that simulations are conducted in the NVE ensemble, which is straightforward to simulate. However, when aiming to understand the current-bias characteristics, especially in the context of the diode effect, the method becomes less convenient. This additional complexity arises due to the need to pre-estimate the temperature drop corresponding to a given current
\cite{JW-HeatMD}.

\section{Results: monomer, dimer and trimer fullerene junctions}
\label{results-sec}

We adopt the three molecular dynamics approaches, the AEMD, LNEMD and RNEMD, and simulate the thermal conductance of fullerene monomer, dimer and trimer junctions.  Overall, we find that the three methods agree in their predictions. Particularly, the RNEMD and the LNEMD are similar in computational effort and in their results. Since the LNEMD is more intuitive to 
operate (one sets up the temperature at the boundaries and generates a heat current in the system), we overall find it to be the most convenient tool among the three.   We now describe our results.

\begin{figure}[h]
    \centering
    \includegraphics[width=\columnwidth]{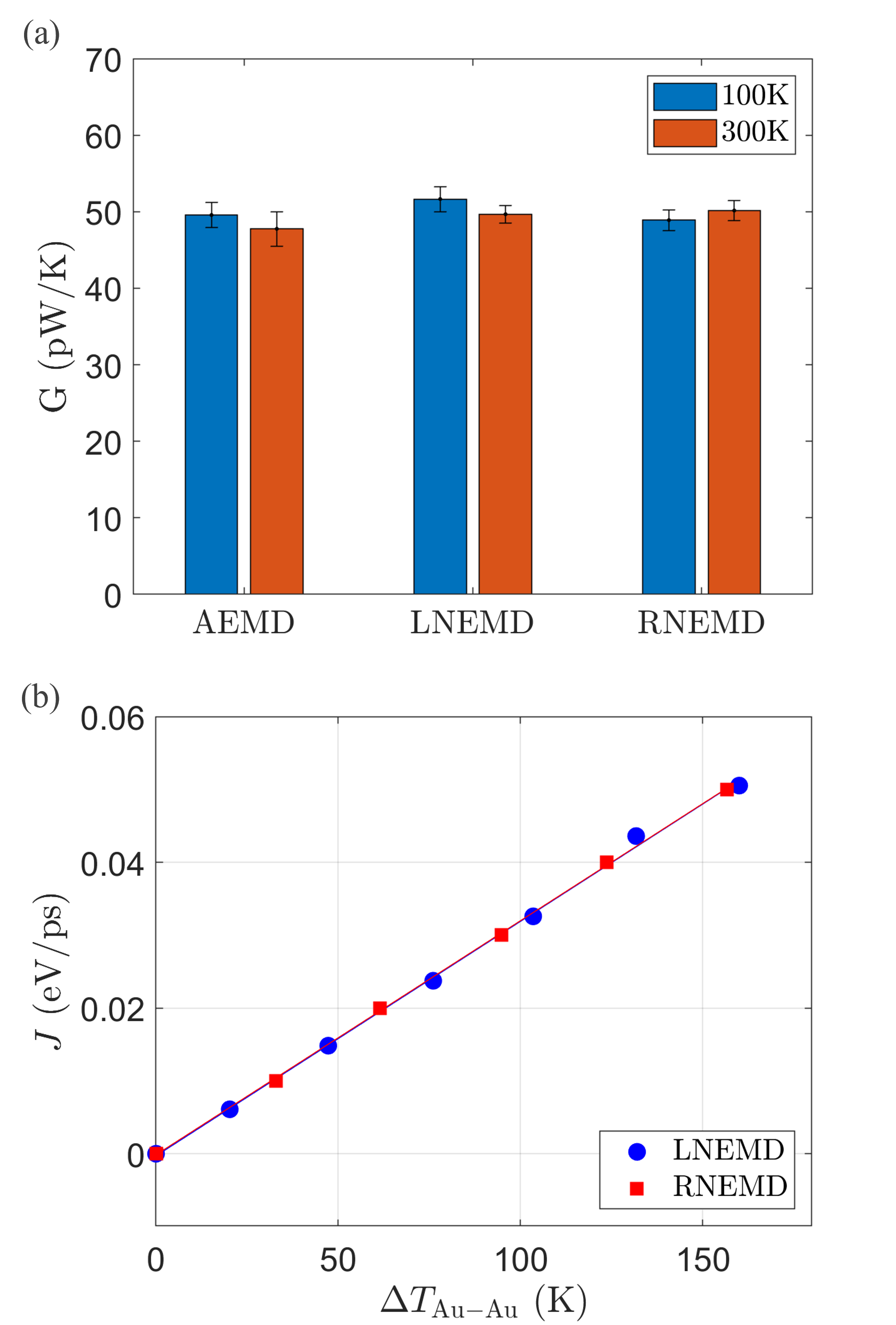}
    \caption{Thermal transport in monomer fullerene junctions. (a) Thermal conductance using different methods as indicated at  $\bar T$=100 K (blue) and $\bar T$=300 K (orange). (b) Heat current as a function of temperature bias in the RNEMD and the LNEMD methods at $\bar T$=300 K.
    We repeated simulations for $\Delta T=100$ K and found that uncertainties in the heat current were smaller than the size of the data point symbol. } 
    \label{Fig4}
\end{figure}

In Fig. \ref{Fig4}(a) we display the thermal conductance of a C60 monomer nanojunction using the three different techniques 
at two different averaged temperatures,
 100 K and 300 K. We make the following observations:
 (i) The thermal conductance of fullerene monomer gold nanojunction does not depend on temperature in the examined range of $\bar T$=100-300 K.
 (ii) The three methods agree, providing thermal conductance of $G\approx 50$ pW/K (49.62 pW/K). 
(iii) Repeating simulations showed small errors of less than 3 pW/K. 
(iv)
In Fig. \ref{Fig4}(b),
we furthermore study the behavior of the heat current as a function of the 
temperature difference, 
 $\Delta T_{\text {Au-Au}}$, and observe a linear trend.

\begin{figure}[htbp]
 \centering
\includegraphics[width=\columnwidth]{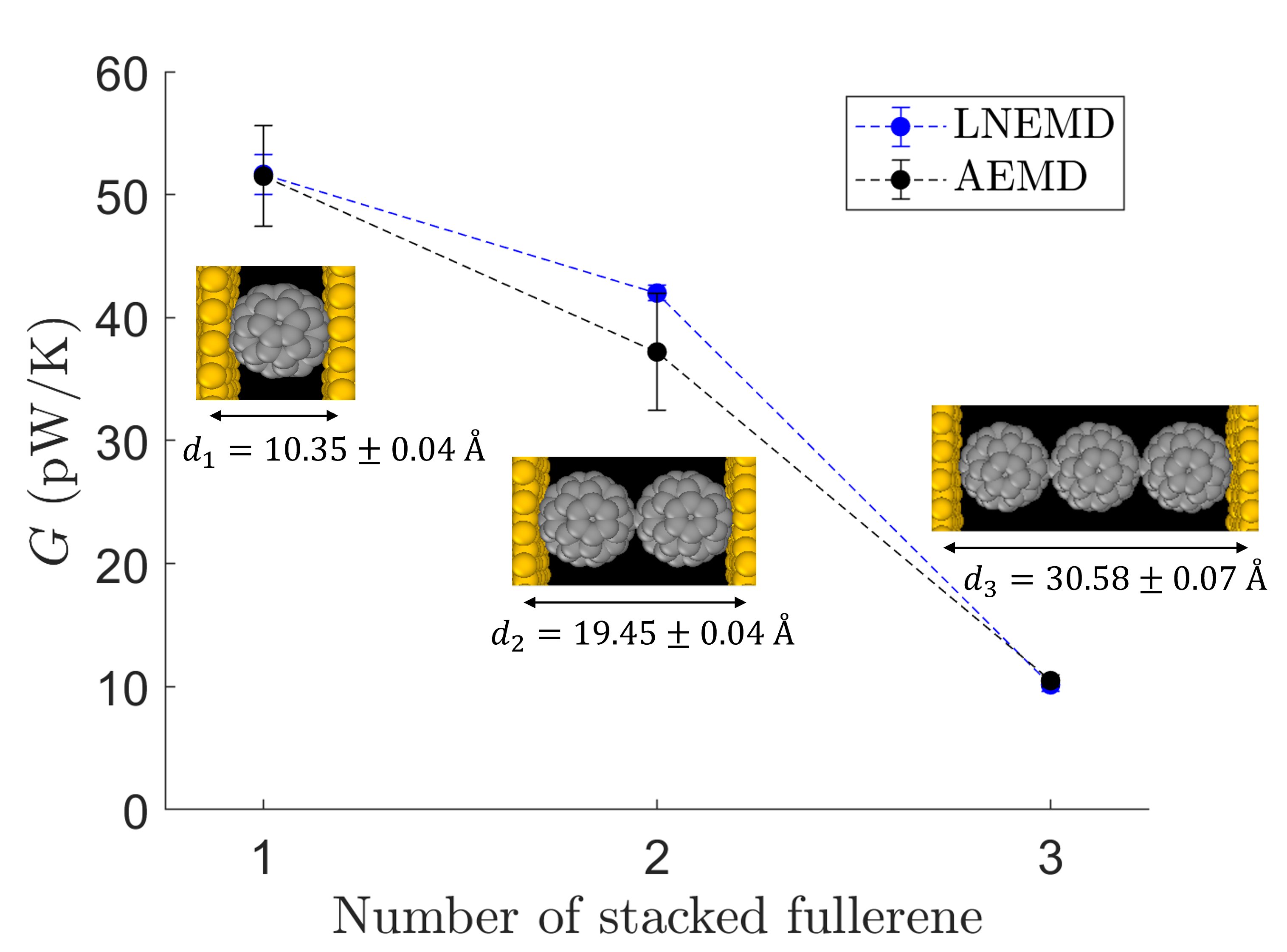}
    \caption{Thermal conductance of monomer, dimer and trimer fullerene junctions as depicted in the illustrations. The length of the metal-to-metal gap is further indicated; deviations during simulations were less than 1\%.
    We compare LNEMD simulations to AEMD results showing a good agreement. 
    }
    \label{Fig5}
\end{figure}

In Fig. \ref{Fig5} we display the thermal conductance of stacked fullerene junctions with $N=1,2,3$ units. We present results using two methods, the AEMD and the LNEMD, and results agree within the uncertainty of simulations.
Significantly, we find that the thermal conductance of fullerene monomer and dimer junctions are comparable, at approximately 50 pW/K and 40 pW/K, respectively. Fullerene trimers however exhibit significantly lower conductance, measured at 10 pW/K. 

\begin{figure}[htbp]
    \centering
\includegraphics[width=\columnwidth]{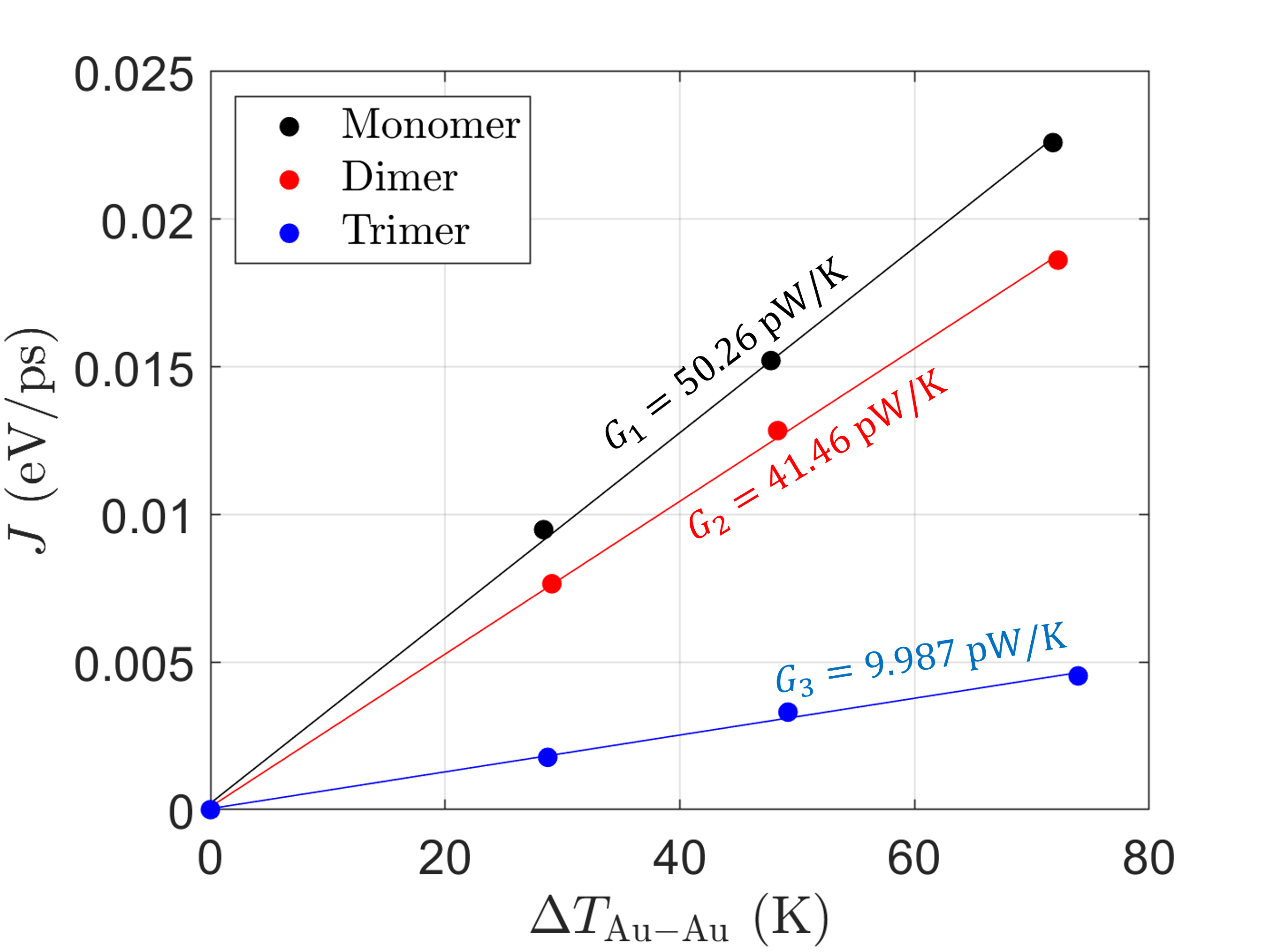}
    \caption{Heat current in stacked fullerenes as a function of temperature difference with $N=1,2,3$ fullerenes top to bottom. The average temperature is 100 K. The slope of each curve, corresponding to the thermal conductance, $G_N$, is indicated in each case.} 
    \label{Fig6}
\end{figure}

Next, in Fig. \ref{Fig6} we display the heat current as a function of the applied temperature difference using the LNEMD method. We also calculate the slope, which corresponds to the thermal conductance. 
We find that even far from equilibrium, for $\Delta T/\bar{T}\approx 1$, stacked fullerene junctions display linear heat current-temperature bias characteristics. 

As for the size of the gap $d$ between the metals, as one can see in the illustration of Fig. \ref{Fig5}, the fullerenes lie in close proximity to each other (efforts to space them out led to their relaxation to this distance); the values $d_1$ and $d_2$ for the monomer and dimers, respectively, are in accord with those tabulated in Ref. \citenum{Klockner}.

The choice of method in future studies depends on specific research objectives. The AEMD has the clear advantage with its shorter simulation time for extracting the thermal conductance. However, as we also discussed in Ref. \citenum{JW-HeatMD}, Eq. (\ref{eq:DTsol}) relies on substantial assumptions, most importantly that a single exponential decay characterizes the equilibration dynamics. 

\begin{figure*}[htbp]
    \centering
\includegraphics[width=2\columnwidth]{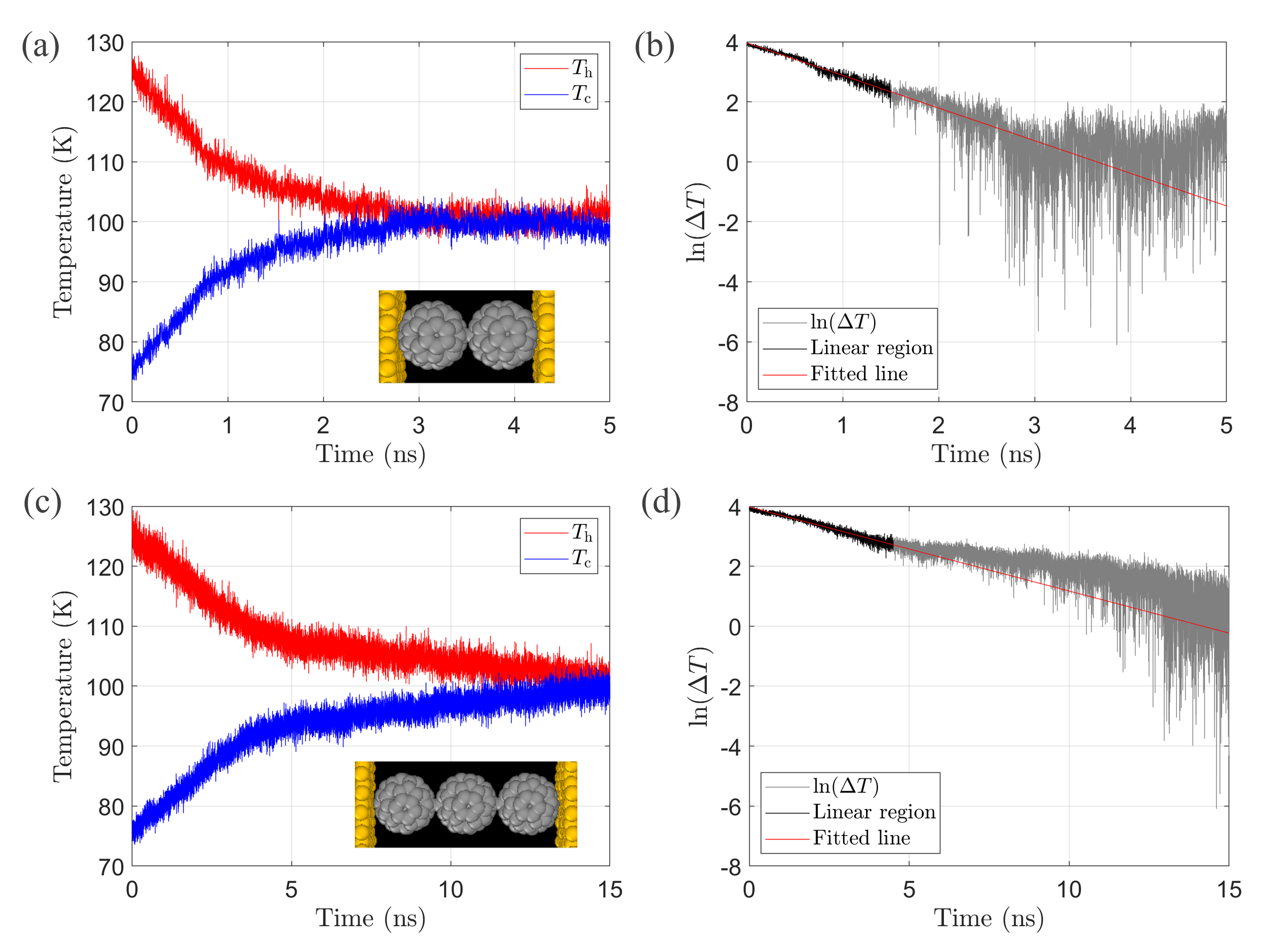}
    \caption{AEMD simulations of (a)-(b) dimer and (c)-(d) trimer fullerene nanojunctions.
    In both cases, we display the temperature of the metals as a function of time, and the corresponding analysis to extract the thermal conductance.}
    \label{Fig7}
\end{figure*}

To assess this assumption, we present in Fig. \ref{Fig7} equilibration traces of the AEMD  method for dimer and trimer junctions, complementing Fig. \ref{Fig2}.
Interestingly, while the dimer fullerene seems to follow a single exponential decay, the equilibration process of a trimer nanojunction is clearly more involved, manifesting two timescales: A relatively fast equilibration (yet slower than in the monomer and dimer junctions), followed by a slower decay. 
We note that we extracted the thermal conductance of Fig. \ref{Fig5} from the initial fast decay, indicated in black in Figs. \ref{Fig7}(b) and (d). 
The AEMD method offers nontrivial information on the equilibration dynamics, yet for thermal conductance calculations we suggest the LNEMD and RNEMD steady-state methods. These methods do not make any assumptions on the underlying dynamics and are thus more general and robust.

We now compare our results to other studies.
In Ref. \citenum{Klockner}, the authors computed the phononic contribution to the thermal conductance quantum mechanically, using the DFT-NEGF method, which however is limited to harmonic interactions. Furthermore, in their calculations, the junction was held at a fixed configuration. This is to be contrasted by our fully classical-dynamical simulations that yet include anharmonic interactions and long-range Lennard-Jones interactions. 
Another noteworthy difference between Ref. \citenum{Klockner} and our work is the geometry of the metal contact. While here we form a junction between two flat metal surfaces, the junction was previously assumed to be formed between two tips, which can be sharp or blunt.

In Table \ref{Table2}, we present results from Ref. \citenum{Klockner} focusing on junctions with two blunt tips. This study further demonstrated that for monomers, the conductance was halved if one of the contacts was made sharp. 
The Table also summarizes our findings from Fig. \ref{Fig6}. Interestingly, both quantum \cite{Klockner} and our classical simulations of monomer fullerene yield similar values for the phononic thermal conductance, at $\approx 45-50$ pW/K, with the classical MD bringing results that are about 10\% higher than the quantum ones, a typical observation as classical simulations often overestimate thermal transport.

DFT-NEGF and MD results though disagree on dimer junctions. While DFT-NEGF calculations provided a low thermal conductance at 7.0 pW/K \cite{Klockner}, our simulations predict that dimers conduct with a similar efficacy to monomers with a thermal conductance that is about 20\% smaller than the monomer, at $\approx$ 40 pW/K. 
The strong suppression of thermal transport shows up only in the trimer junction, where MD predicts thermal conductance at $\approx$ 10 pW/K (we do not have results from the literature to compare to this value).
We analyzed the temperature profiles of dimer and trimer fullerenes, similar to the monomer data presented in Fig. \ref{Fig3}. 
We observed (not shown) a very small temperature drop between the fullerenes, at the order of  0.5 K 
for the dimer (and smaller for the trimer).
Thus, internal contact resistance between the fullerenes cannot explain the substantial reduction in thermal conductance for trimer junctions.

\begin{figure}[htbp]
    \centering
\includegraphics[width=\columnwidth]{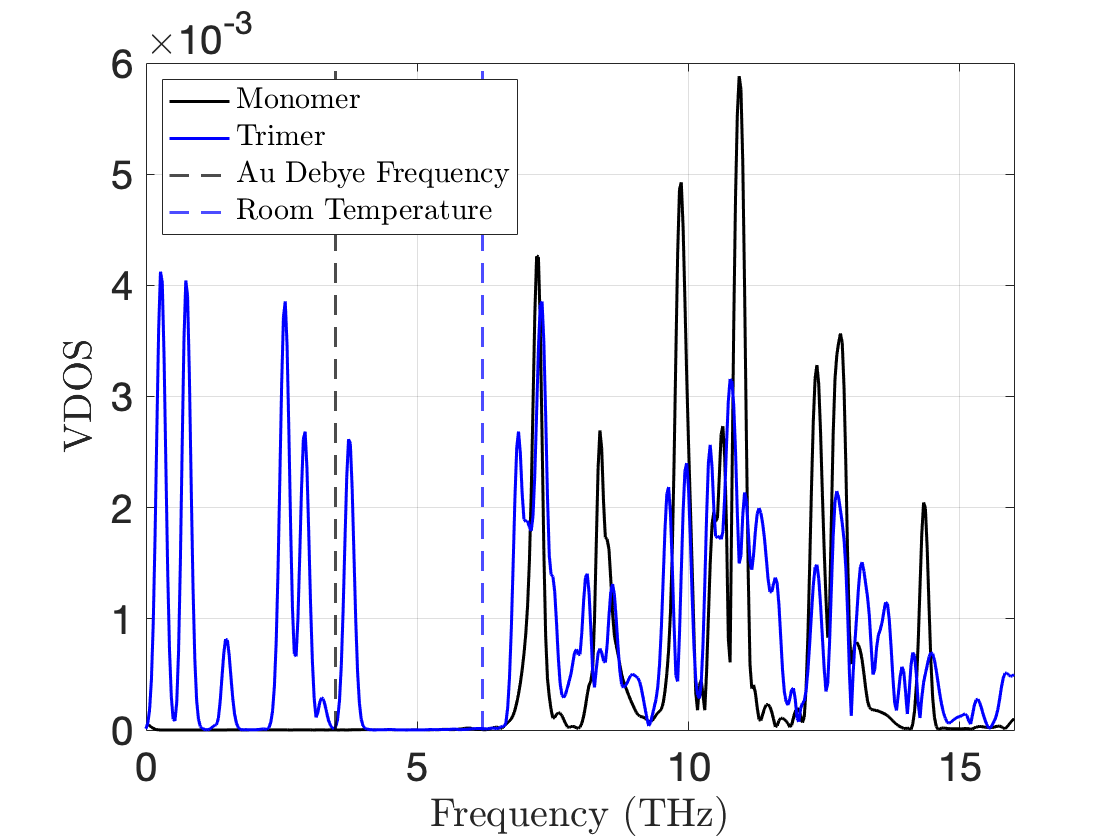}
    \caption{Vibrational density of states (VDOS) for monomer and trimer fullerene as a function of frequency. The average temperature is 300 K. The black dashed line roughly marks the Debye frequency of gold, while the dashed blue line stands at room temperature. } 
    \label{Fig8}
\end{figure}

To address the question of why there is a large difference in thermal conductances, we plot in Fig. \ref{Fig8} the vibrational density of states (VDOS) for isolated fullerenes based on velocity data from LAMMPS \cite{Bringuier, VDOS}. Below 6 THz, which is roughly the Debye frequency of gold, we anticipated the monomer VDOS to exhibit more peaks than the trimer's.
In contrast, the trimer's VDOS displays more structure at low frequencies, likely representing modes associated with inter fullerene motion. However, these modes do not appear to contribute to the metal-to-metal thermal transport. The suppression of thermal conductance for trimer junctions thus remains an open question.

Going back to Table II, we comment that other experimental and computational studies on the thermal transport properties of fullerenes have explored a variety of configurations, rather than the metal-molecule-metal junction considered in our study. 
The wide variety of studied systems include fullerenes in solution \cite{Keblinski-AEMD,vib}, crystalline fullerenes\cite{Kumar15}, fullerenes with molecules confined within them \cite{Gao}, and fullerenes manufactured into  materials\cite{PRAM,Zhao24}.
Furthermore, these studies typically focused on  bulk thermal conductivity rather than thermal conductance, reflecting the consideration of fullerenes as a carbon-based nanomaterial. 
In the above mentioned systems, reported thermal conductivity for C$_{60}$ ranged from 0.1-0.4 W/m K\cite{vib,Kumar15,PRAM,Zhao24,Cahill13}.
 %

\begin{table}[hbtp]
	\begin{center}
	\caption{Table of computed thermal conductance of fullerene junctions}
	\footnotesize
	\begin{tabular}{|m{6em}|m{4.5em}|m{5em}|m{5.5em}|m{4.5em}|m{4.5em}|}
		\hline
		\bf  Fullerene configuration & \bf Metal geometry & \bf Separation  ($d$) & \bf Conductance (pW/K))& \bf Method & \bf Reference \\
		\hline
(i) Single & Blunt tips & 1.11 nm &  46.3 & DFT-NEGF &Ref. \cite{Klockner} \\
		\hline
(ii) Dimer & Blunt tips & 1.91 nm & 7.0 & DFT-NEGF& Ref. \cite{Klockner} \\
		\hline
(iii) Single & Flat metal & 1.04 nm & 50.3  &LNEMD &Fig. \ref{Fig6} \\
		\hline
(iv) Dimer & Flat metal & 1.95 nm & 41.5 & LNEMD  &Fig. \ref{Fig6} \\
  		\hline
(v) Trimer & Flat metal &3.06 nm & 10.0 & LNEMD  &Fig. \ref{Fig6} \\
		\hline
	\end{tabular}
\label{Table2}
\end{center}
\end{table}
%

\section{Discussion and Summary}
\label{Sum-sec}

In this study, we investigated phononic thermal transport in single-molecule gold-fullerene-gold nanojunctions based on classical molecular dynamics simulations.
We computed the thermal conductance of monomer, dimer and trimer fullerene molecules sandwiched between metals. We showed that the thermal conductance decreased monotonically with the number of fullerene units. Notably, the thermal conductance of trimer junctions was five times smaller than that of monomer junctions, and four times smaller than dimer junctions. 

Our study further affirmed the consistency of different classical MD tools in studying thermal transport properties of nanojunctions. 
Particularly, we showed that the AEMD method, which extracts the thermal conductance values from transient dynamics, recovered correct results, and even when the relaxation dynamics followed a multi-exponential decay form.
The next stage of such simulations would be to employ first principle-based parametrization for the atomic potentials, and the inclusion of quantum effects through corrections to the thermostats, as done in Ref. \citenum{Lu2021}. These extensions require understanding of the underlying MD method, thus consistency checks performed here serve as a foundation for such more sophisticated modelling.

There are a plethora of open questions and simulation geometries that could be explored in future work. For example, it was demonstrated in ref. \citenum{Klockner} that the metal-to-metal separation strongly affects the heat current. In the present work, we could not examine this aspect as the bulk metal electrodes tended to approach each other upon equilibration. One could however enforce a certain metal-to-metal gap by fixing (making immobile) a large portion of atoms at the boundaries.
Additionally, the structure of the tip plays a significant role in thermal transport as illustrated in Ref. \citenum{Klockner}. This aspect was not examined in our work and we studied only the case of fullerenes placed between flat surfaces (which could represent very blunt tips).
Finally, we contained ourselves here with gold metals, but results should depend on the type of the contact, particularly when using graphene electrodes instead of gold \cite{Lu2021}. Other interesting extensions concern thermal transport characteristics of endohedral fullerenes, using encloses atoms or molecules as knobs to control thermal transport  \cite{Gao}.
It is our hope that this work will spur additional computational and theoretical studies of heat transport in single-molecule junctions, and motivate new experimental efforts in this field.

\begin{acknowledgements}
DS acknowledges support from an NSERC Discovery Grant and the Canada Research Chair program. The work of JL was supported by an NSERC USRA award. 
\end{acknowledgements}


\begin{thebibliography}{99}


\bibitem{CuiExp19}
L. Cui, S. Hur, Z. A. Akbar, J. C. Kl\"ockner, W. Jeong, F. Pauly, S.-Y. Jang, P. Reddy, and E. Meyhofer,
"Thermal conductance of single-molecule junctions,"
\href{https://doi.org/10.1038/s41586-019-1420-z}{Nature} {\bf 572}, 628 (2019).

\bibitem{GotsmannExp19}
N. Mosso, H. Sadeghi, A. Gemma, S. Sangtarash, U. Drechsler, C. Lambert, and B. Gotsmann,
"Thermal Transport through Single-Molecule Junctions,"
\href{https://doi.org/10.1021/acs.nanolett.9b02089}{Nano Lett.} {\bf 19}, 7614 (2019).
%

\bibitem{GotsmannExp23}
A. Gemma, F. Tabatabaei, U. Drechsler, et al., 
"Full thermoelectric characterization of a single molecule,"
\href{https://doi.org/10.1038/s41467-023-39368-7}{Nat. Commun.} {\bf 14}, 3868 (2023).

\bibitem{Baowen12}
N. Li, J. Ren, L. Wang, G. Zhang, P. H\"anggi, and B. Li,
"Colloquium: Phononics: Manipulating heat flow with electronic analogs and beyond,"
\href{https://doi.org/10.1103/RevModPhys.84.1045}{Rev. Mod. Phys.} {\bf 84}, 1045 (2012).

\bibitem{Leitner15}
D. M. Leitner, 
"Quantum Ergodicity and Energy Flow in Molecules,"
\href{https://doi.org/10.1080/00018732.2015.1109817}{Adv. Phys.} {\bf 64}, 445 (2015).

\bibitem{RevA}
D. Segal and  B. K. Agarwalla,
"Vibrational Heat Transport in Molecular Junctions,"
\href{https://doi.org/10.1146/annurev-physchem-040215-112103}{Ann. Rev. Phys. Chem.} {\bf 67}, 185 (2016).

\bibitem{RevG}
B. Gotsmann, A. Gemma, and D. Segal,
"Quantum phonon transport through channels and molecules—A Perspective,"
\href{https://doi.org/10.1063/5.0088460}{App. Phys. Lett.} {\bf 120}, 160503 (2022).

\bibitem{Pop10}
E. Pop,
"Energy dissipation and transport in nanoscale devices,"
\href{https://doi.org/10.1007/s12274-010-1019-z}{Nano Res.} {\bf 3}, 147 (2010).

\bibitem{Rev14}
D. G. Cahill,  P. V. Braun, G. Chen, D. R. Clarke, S. Fan, K. E. Goodson, P. Keblinski, W. P. King, G. D. Mahan, A. Majumdar, H. J. Maris, S. R. Phillpot, E. Pop, and L. Shi,
"Nanoscale thermal transport. II. 2003–2012,"
\href{https://doi.org/10.1063/1.4832615}{Appl. Phys. Rev.} {\bf 1}, 011305 (2014).

\bibitem{PramodR}
K. Wang, E. Meyhofer, and P. Reddy,
Thermal and Thermoelectric Properties of Molecular Junctions,
\href{https://doi.org/10.1002/adfm.201904534}{Advanced Functional Materials} {\bf 30}, 1904534 (2019).

\bibitem{Yoon20}
S. Park, J. Jang, H. Kim, D. I. Park, K. Kim, and H. J. Yoon,
"Thermal conductance in single molecules and self-assembled monolayers: physicochemical insights, progress, and challenges,"
\href{https://doi.org/10.1039/D0TA07095E}{J. Mater. Chem. A} {\bf 8}, 19746 (2020).

\bibitem{BaowenR21}
Y. Li, W. Li, T. Han, X. Zheng, J. Li, B. Li, S. Fan, and C.-W. Qiu,
"Transforming heat transfer with thermal metamaterials and devices,"
\href{https://doi.org/10.1038/s41578-021-00283-2}{Nat. Rev. Mater.} {\bf 6}, 488 (2021).


\bibitem{alkaneR}
N. J. Tao, 
Electron transport in molecular junctions,
Nature {\bf 1}, 173 (2006).



\bibitem{Wang06}
R. Y. Wang, R. A. Segalman, and A. Majumdar,
"Room temperature thermal conductance of alkanedithiol self-assembled monolayers,"
\href{https://doi.org/10.1063/1.2358856}{Appl. Phys. Lett.} {\bf 89}, 173113 (2006).

\bibitem{Dlott07}
Z. Wang, J. A. Carter, A. Lagutchev, Y. K. Koh, N.-H. Seong, D. G. Cahill, and D. D. Dlott,
"Ultrafast Flash Thermal Conductance of Molecular Chains,"
\href{https://doi.org/10.1126/science.1145220}{Science} {\bf 317}, 787 (2007).

\bibitem{Cahill12}
M. D. Losego, M. E. Grady, N. R. Sottos, D. G. Cahill, and P. V. Braun,
"Effects of chemical bonding on heat transport across interfaces,"
\href{https://doi.org/10.1038/nmat3303}{Nat. Mater.} {\bf 11}, 502 (2012).

\bibitem{GotsmannExp14}
T. Meier, F. Menges, P. Nirmalraj, H. H\"olscher, H. Riel, and B. Gotsmann,
"Length-Dependent Thermal Transport along Molecular Chains,"
\href{https://doi.org/10.1103/PhysRevLett.113.060801}{Phys. Rev. Lett.} {\bf 113}, 060801 (2014).


\bibitem{Shub15}
S. Majumdar, J. A. Sierra-Suarez, S. N. Schiffres, W.-L. Ong, C. F. Higgs III, A. J. H. McGaughey, and J. A. Malen,
"Vibrational Mismatch of Metal Leads Controls Thermal Conductance of Self-Assembled Monolayer Junctions,"
\href{https://doi.org/10.1021/nl504844d}{Nano Lett.} {\bf 15}, 2985 (2015).

\bibitem{Shub17}
S. Majumdar, J. A. Malen, and A. J. H. McGaughey,
"Cooperative Molecular Behavior Enhances the Thermal Conductance of Binary Self-Assembled Monolayer Junctions,"
\href{https://doi.org/10.1021/acs.nanolett.6b03894}{Nano Lett.} {\bf 17}, 220 (2017).



\bibitem{Dvira2003}
D. Segal, A. Nitzan, and P. H\"anggi,
"Thermal conductance through molecular wires,"
\href{https://doi.org/10.1063/1.1603211}{J. Chem. Phys.} {\bf 119}, 6840 (2003).

\bibitem{Pawel11}
K. Sasikumar and P. Keblinski,
"Effect of chain conformation in the phonon
transport across a Si-polyethylene single-
molecule covalent junction,"
\href{https://doi.org/10.1063/1.3592296}{J. Appl. Phys.} {\bf 109}, 114307 (2011). 

\bibitem{Pauly16}
J. C. Kl\"ockner, M. B\"urkle, J. C. Cuevas, and F. Pauly,
"Length dependence of the thermal conductance of alkane-based single-molecule junctions: An ab initio study,"
\href{https://doi.org/10.1103/PhysRevB.94.205425}{Phys. Rev. B} {\bf 94}, 205425 (2016).

\bibitem{Pauly18}
J. C. Kl\"ockner, J. C. Cuevas, and F. Pauly,
"Transmission eigenchannels for coherent phonon transport,"
\href{https://doi.org/10.1103/PhysRevB.97.155432}{Phys. Rev. B} {\bf 97}, 155432 (2018).


\bibitem{Nitzan20}
I. Sharony, R. Chen, and A. Nitzan,
"Stochastic simulation of nonequilibrium heat conduction in extended molecular junctions,"
\href{https://doi.org/10.1063/5.0022423}{J. Chem. Phys.} {\bf 153}, 144113 (2020).

\bibitem{Lu2021}
G. Li, B.-Z. Hu, N. Yang, and J.-T. L\"u,
"Temperature-dependent thermal transport of single molecular junctions from semiclassical Langevin molecular dynamics,"
\href{https://doi.org/10.1103/PhysRevB.104.245413}{Phys. Rev. B} {\bf 104}, 245413 (2021).

\bibitem{Nitzan22}
M. Dinpajooh and A. Nitzan,
"Heat conduction in polymer chains: Effect of substrate on the thermal conductance,"
\href{https://doi.org/10.1063/5.0087163}{J. Chem. Phys.} {\bf 156}, 144901 (2022).

\bibitem{JW-HeatMD}
J. J. Wang, J. Gong, A. J. H. McGaughey, and D. Segal, "Simulations of heat transport in single-molecule junctions: Investigations of the thermal diode effect,” \href{https://doi.org/10.1063/5.0125714}{J. Chem. Phys.} {\bf 157}, 174105 (2022).

\bibitem{JW-fluct}
J. J. Wang, M. Gerry, and D. Segal,
"Challenges in molecular dynamics simulations of heat exchange statistics,"
\href{
https://doi.org/10.48550/arXiv.2311.07830
}{arXiv:2311.07830}.

\bibitem{Luo13}
T. Luo and G. Chen, 
"Nanoscale Heat Transfer - from Computation to Experiment," 
\href{https://doi.org/10.1039/C2CP43771F}{Phys. Chem. Chem. Phys.} {\bf 15}, 3389 (2013).




\bibitem{DFTNEGF1}
M. Bürkle, T. J. Hellmuth, F. Pauly, and Y. Asai, 
"First-principles calculation of the thermoelectric figure of merit for [2,2]paracyclophane-based single-molecule junctions,"
\href{https://doi.org/10.1103/PhysRevB.91.165419}{Phys. Rev. B} {\bf 91}, 165419 (2015).

\bibitem{LAMMPS}
A. P. Thompson, H. M. Aktulga, R. Berger, D. S. Bolintineanu, W. M. Brown, P. S. Crozier, P. J. in 't Veld, A. Kohlmeyer, S. G. Moore, T. D. Nguyen, R. Shan, M. J. Stevens, J. Tranchida, C. Trott, and S. J. Plimpton,
"LAMMPS - a flexible simulation tool for particle-based materials modeling at the atomic, meso, and continuum scales,"
\href{https://doi.org/10.1016/j.cpc.2021.108171}{Comp. Phys. Comm.} {\bf 271}, 10817 (2022).










\bibitem{Fullerene-rev}
E. Ghavanloo, H. Rafii-Tabar, A. Kausar, G. I. Giannopoulos, and S. A. Fazelzadeh, 
"Experimental and computational physics of fullerenes and their nanocomposites: Synthesis, thermo-mechanical characteristics and nanomedicine applications," \href{https://doi.org/10.1016/j.physrep.2022.10.003}{Phys. Rep.} {\bf 996}, 1 (2023).

\bibitem{Evangeli} 
C. Evangeli, K. Gillemot, E. Leary, M. T. González, G. Rubio-Bollinger, C. J. Lambert, and N. Agra\"{i}t, "Engineering thermopower of C60 molecular junctions," \href{https://doi.org/10.1021/nl400579g}{Nano Lett.} {\bf 13}, 2141 (2013).

\bibitem{Klockner} 
J. C. Kl\"{o}ckner, R. Siebler, J. C. Cuevas, and F. Pauly, "Thermal conductance and thermoelectric figure of merit of C60-based single-molecule junctions: Electrons, phonons, and photons," \href{https://doi.org/10.1103/PhysRevB.95.245404}{Phys. Rev. B} {\bf 95}, 245404 (2017).

\bibitem{OVITO}
A. Stukowski, "Visualization and analysis of atomistic simulation data with OVITO-the Open Visualization Tool," \href{https://iopscience.iop.org/article/10.1088/0965-0393/18/1/015012}{Modelling Simul. Mater. Sci. Eng.} {\bf 18}, 015012 (2010).

\bibitem{AIREBO}
S. J. Stuart, A. B. Tutein,  and J. A. Harrison, 
``A reactive potential for hydrocarbons with intermolecular interactions,” \href{https://doi.org/10.1063/1.481208}{J. Chem. Phys.} {\bf 112} (14), 6472–6486 (2000).

\bibitem{Pishkenari} 
H. N. Pishkenari, A. Nemati, A. Meghdari, and S. Sohrabpour, "A close look at the motion of C60 on gold," \href{https://doi.org/10.1016/j.cap.2015.08.003}{Curr. Appl. Phys.}  {\bf 15}, 1402 (2015).

\bibitem{Grochola}
G. Grochola, S. P. Russo, and I. K. Snook, "On fitting a gold embedded atom method potential using the force matching method," \href{https://doi.org/10.1063/1.2124667}{J. Chem. Phys.} {\bf 123}, 204719 (2005).

\bibitem{Keblinski-AEMD}
 S. T. Huxtable, D. G. Cahill, S.  Shenogin, P. Keblinski,
 "Relaxation of Vibrational Energy in Fullerene Suspensions," 
\href{https://www.sciencedirect.com/science/article/pii/S000926140500391X?via%3Dihub}
{Chem. Phys. Lett. }{\bf 407}, 129 (2005).


\bibitem{Martin}
T.-Q. Duong, C. Massobrio, G. Ori, M. Boero, and E. Martin, 
"Thermal
resistance of an interfacial molecular layer by first-principles molecular dynamics," \href{https://doi.org/10.1063/5.0014232}{J. Chem. Phys.} {\bf 153}, 074704 (2020).


\bibitem{ITR}
J. Chen, X. Xu, J. Zhou, and B. Li,
"Interfacial thermal resistance: Past, present, and future,"
\href{https://journals.aps.org/rmp/abstract/10.1103/RevModPhys.94.025002}
{Rev. Mod. Phys.} {\bf 94}, 025002 (2022).

\bibitem{ITR2}
A. Giri, S. G. Walton, J. Tomko, N. Bhatt, M. J. Johnson, D. R. Boris, G. Lu, J. D. Caldwell, O. V. Prezhdo, and P. E. Hopkins
"Ultrafast and Nanoscale Energy Transduction Mechanisms and Coupled Thermal Transport across Interfaces," 
\href{https://doi.org/10.1021/acsnano.3c02417}
{ACS Nano} {\bf 17}, 14253 (2023). 

\bibitem{Lloyd}
T. Luo and J. R. Lloyd, "Non-equilibrium molecular dynamics study of
thermal energy transport in Au–SAM–Au junctions," \href{https://doi.org/10.1016/j.ijheatmasstransfer.2009.10.033}{J. Heat Mass Transf.} 53, 1 (2010).

\bibitem{Keblinski}
L. Hu, L. Zhang, M. Hu, J.-S. Wang, B. Li, and P. Keblinski, "Phonon
interference at self-assembled monolayer interfaces: Molecular dynamics simulations," \href{https://doi.org/10.1103/PhysRevB.81.235427
}{Phys. Rev. B} {\bf 81}, 235427 (2010).


\bibitem{Kiku} G. Kikugawa, T. Ohara, T. Kawaguchi, I. Kinefuchi, and Y. Matsumoto,
"A molecular dynamics study on heat conduction characteristics inside the alkanethiolate SAM and alkane liquid," \href{https://doi.org/10.1016/j.ijheatmasstransfer.2014.07.040}{J. Heat Mass Transf.} {\bf 78}, 630 (2014).

\bibitem{Chen}
Z. Li, S. Xiong, C. Sievers, Y. Hu, Z. Fan, N. Wei, H. Bao, S. Chen,
D. Donadio, and T. Ala-Nissila, "Influence of thermostatting on nonequilibrium molecular dynamics simulations of heat conduction in solids," \href{https://doi.org/10.1063/1.5132543}{J. Chem. Phys.} {\bf 151}, 234105 (2019).







\bibitem{Bringuier}
S. Bringuier, dump2VDOS.py: Vibrational density of states from LAMMPS dump file (2.1), \href{https://doi.org/10.5281/zenodo.10573320}{Zenodo} (2024).

\bibitem{VDOS}
S. Gonçalves and H. Bonadeo, "Vibrational densities of states from molecular-dynamics calculations," \href{http://doi.org/10.1103/PhysRevB.46.12019}{Phys. Rev. B} {\bf 46}, 12019 (1992).



\bibitem{vib}
C. J. Szwejkowski, A. Giri†, R. Warzoha, B. F. Donovan, B. Kaehr, and P. E. Hopkins,
"Molecular Tuning of the Vibrational Thermal Transport Mechanisms in Fullerene Derivative Solutions,"
\href{https://pubs.acs.org/doi/epdf/10.1021/acsnano.6b06499}
{ACS Nano} {\bf 11}, 1389 (2017).


 \bibitem{Kumar15}
L. Chen, X. Wang, and S. Kumar,
"Thermal Transport in Fullerene Derivatives Using Molecular Dynamics Simulations,"
\href{https://www.nature.com/articles/srep12763}{Scientific Reports} {\bf 5}, 12763 (2015).

\bibitem{Gao} 
Y. Gao and B. Xu, "Probing Thermal Conductivity of Fullerene C60 Hosting a Single Water Molecule," \href{https://doi.org/10.1021/acs.jpcc.5b05663}{J. Phys. Chem. C} {\bf 119}, 20466 (2015).

\bibitem{PRAM}
C. Kim, D.-S. Suh, K. H. P. Kim, Y.-S. Kang, T.-Y. Lee, Y. Khang, and D. G. Cahill, "Fullerene thermal insulation for phase change memory," \href{https://doi.org/10.1063/1.2830002}{Appl. Phys. Lett.} {\bf 92}, 013109 (2008).

\bibitem{Zhao24}
Z. Li, Y. Chen, Z.-H. Li, Y. Zhang, N. Wei, Y. Cheng, and J. Zhao, "Thermal Property of Fullerene Fibers: One-Dimensional Material with Exceptional Thermal Performance," \href{https://doi.org/10.1002/smll.202307671}{Small} {\bf 2024}, 2307671 (2024).

\bibitem{Cahill13}
X. Wang, C. D. Liman, N. D. Treat, M. L. Chabinyc, and D. G. Cahill, "Ultralow thermal conductivity of fullerene derivatives," \href{http://dx.doi.org/10.1103/PhysRevB.88.075310}{Phys. Rev. B} {\bf 88}, 075310 (2013).




\end{thebibliography}
\end{document}